\documentclass[12pt,a4paper]{article}
\usepackage{amsmath,amsthm,amsfonts,amssymb,alltt}
\usepackage{graphics}
\usepackage{graphicx}
\usepackage{epsfig}

\theoremstyle{definition}

\theoremstyle{remark}

\numberwithin{equation}{section}

\begin{document}

\author{Simon Gluzman, Vladimir Mityushev
\\ Dept. Computer Sciences and Computer Methods, 
\\ Pedagogical University, \\ ul. Podchorazych 2, Krakow 30-084, Poland}

\title{Series, Index and Threshold for Random 2D Composite} 
\date{}
\maketitle
\begin{abstract}
Effective conductivity of a 2D random composite is expressed in
the form of long series in the volume fraction of ideally conducting
disks. The problem of a {\it direct} reconstruction of the critical index for
superconductivity from the series is solved with good accuracy, for the
first time. General analytical expressions for conductivity in the whole range of concentrations are derived and compared with the regular composite
and existing models.
\end{abstract}

\section{Introduction} 
It is frequently declared that only lower order formulae can be deduced for the effective conductivity problem which cannot be analytically solved in general case because of the complicated random geometrical structures. After such an announce hard numerical computations are applied to solve such a problem. Of course, advanced computational approaches can be useful in mechanical engineering. But an exact or approximate analytical formula is always better because it can exactly show asymptotic behavior near singular points when numerics usually fails. 

\begin{figure}
\begin{center}
\includegraphics[scale=0.6]{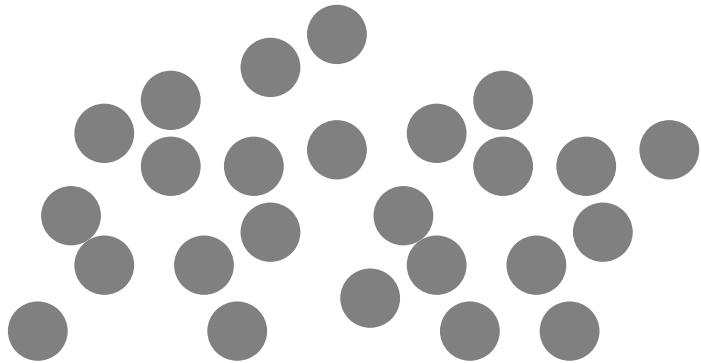}
\end{center}
\caption{
Randomly distributed disks.
}
\label{figDisksRandom}
\end{figure}

In the present paper, we deduce such a formula for a 2D, two-component composite made from a collection of non-overlapping, identical, circular discs, embedded randomly in an otherwise uniform locally isotropic host (see Fig.\ref{figDisksRandom}). 
The conductivity of the host is normalized to unity. The effective conductivity problem for an insulating or ideally conducting inclusions is called the conductivity and superconductivity problem, respectively \cite{torkbook}. The problem and its approximate solution go back to Maxwell, see e.g. \cite{Mit2013b}. 

There are two important unresolved problems in the theory of random composites:

1. what quantity should stand for the maximum volume fraction $x_{c}$ of random composites \cite{torkreview}, and

2.  theoretical explanation of  the values of critical indices for conductivity and superconductivity denoted by $t$ and $s$, respectively \cite{torkbook}.

Recently, a novel technique for deriving expansions in concentration was suggested \cite{method}.
It combines analytic and numeric methods for solving the conductivity problem directly in the 2D case. It is applicable both for regular \cite{ourpaper} and random cases. Thus, we proceed to the case of a 2D random composite, where rather long series in concentration  for the effective conductivity by itself, will be presented and analyzed systematically, following generally to \cite{method, ourpaper}. The series will be  used to estimate the index and  the threshold in 2D random case.

The considered problem can be equivalently formulated as follows. Given the polynomial approximation \eqref{seriesrand} of the function $\sigma(x)$, to estimate the convergence radius $x_c$ of the Taylor series of $\sigma(x)$, and to determine parameters of the asymptotically equivalent approximation\eqref{crit} near $x=x_c$.

The problem of defining the threshold is highly non-trivial, since the random closest packing of hard spheres turned out to be ill-defined, and cannot stand for the maximum volume fraction.  It depends on the protocol employed to produce the random packing as well as other system characteristics \cite{torkreview}.  

The problem seems less acute in two dimensions, where various protocols
seems to agree on what quantity should stand for the maximum volume fraction of random composites \cite{protoc1,protoc2, protoc3, method, CzNawMit2}. Namely it is the concentration of  $\frac{\pi}{\sqrt{12}}\approx 0.9069$, 
attained only for the regular hexagonal array of disks. The sought value for a long time was thought to be close to
$0.82$, and considered as random close packing value \cite{berr}. It was recognized recently, that it does not correspond to the maximally random jammed state \cite{torkreview}. For volume fractions  above $0.82$  some local order is present and irregular packing is polycrystalline, forming
rather large triangular coordination domains-grains. In present paper, a protocol with $x_c=\frac{\pi}{\sqrt{12}}$ is used, although  our method can be applied with another protocol with unknown $x_c$. 

All attempts to explain the value of critical indices through geometrical quantities of percolation problem, i.e. universally \cite{torkbook,chou}, had failed so far and the indices are considered independent.  From the phase interchange theorem \cite{kell0} it follows that in two-dimensions, the superconductivity index is equal to the conductivity index \cite{torkbook}, \cite{ryl}, \cite{ryl2008}. 

While it is clear that using expansions in concentration for the conductivity, one should be able to address the two problems, in practice there are no more than two terms available for random systems \cite{chou}, because of the serious technical difficulties.
No method even such powerful as renormalization, or resummation approaches can draw reliable conclusions systemically, based on such short series \cite{chou}. "In fact, the age-old method of series expansions is also blocked by the same
difficulties... "\cite{chou}.

This concerns also self consistent methods (SCMs) which include Maxwell's approach, effective medium approximations, differential schemes etc. SCMs are valid only for dilute composites when the interactions between inclusions do not matter \cite{Mit2013b}. 
The idea to correct a self consistent method (SCM) result $t=s=1$ in all dimensions remained, therefore, theoretically unattainable (see, nevertheless, \cite{fix}). 

We should also mention an indirect approach to estimating $t$ for resistor networks from resistive susceptibility via scaling relations \cite{low}. This approach also dwells heavily on resummation techniques.

%

\section{Series for Conductivity, Random 2D}
In order to correctly define the effective conductivity tensor $\boldsymbol\sigma$ of random composites, the probabilistic distribution of disks of radius $r$ must be introduced, since already the second order term of $\boldsymbol\sigma$ in concentration depends on the distribution \cite{Mit2013b}.  
For macroscopically isotropic composites, the third order term begins to depend on the distribution \cite{torkbook, Mit2013b}.  In the present paper, we consider the uniform non-overlapping distribution when    
a set of independent and identically distributed (i.i.d.) points $\mathbf a_i$ are located in the plane in such a way that $|\mathbf a_i - \mathbf a_j| \geq 2r$. 

For $r=0$ we arrive at the Poisson distribution and for the maximally possible concentration $x_c=\frac{\pi}{\sqrt{12}}$, the distribution degenerates to the unique location, the hexagonal array. The tensor $\boldsymbol\sigma$ is expressed through the scalar effective conductivity $\sigma$ as follows $\boldsymbol\sigma = \sigma \mathbf I$, where $\mathbf I$ is the unit tensor.
In the present paper, the numerical computations are performed only for the hexagonal representative cell. 

This assumption does not restrict our investigation since the number of
inclusions per cell can be taken arbitrary large, hence, the shape of the cell does not impact on the final result.

Consider sufficiently large number of
non-overlapping circular disks of radius $r$ with
the centers $\mathbf a_k$.
The formal definition of the random variable  has to be statistically realized to get numerical results. The protocol for the data is based on the Monte Carlo simulations \cite{method} and can be shortly described as follows. At the beginning, the centers $\mathbf a_k$ are located at the nodes of the regular hexagonal lattice and further randomly moved without overlapping. After sufficiently long random walks the centers form a statistical event satisfying the considered distribution. Using these locations of disks we compute coefficients of $\sigma$ in $x$ many times and take the average. 
Detailed description of the computational method and all relevant parameters for simulations can be found in \cite{method,CzNawMit2}. The method yields
\begin{equation}
\label{seriesrand}
\begin{array}{llll}
\sigma{(x)}=
1 + 2 x + 2 x^2 +2.765912418226355 x^3 +
\\
 8.485557502521662 x^4+0.8698170539309313 x^5+ 
\\
0.02832722167119779 x^6 + 0.03837167966919316 x^7 +
\\
0.17554742683813554 x^8 + 0.2170785960242611 x^9 +
\\
0.08498671921129161 x^{10} + 0.008233910943750663 x^{11} +
\\
0.380088666905241 x^{12} +1.442357383098656 x^{13} +
\\
3.121524280853671 x^{14} + 5.104077444715624 x^{15} +
\\
7.018649741070781 x^{16}+ 8.580574311676896 x^{17} +
\\
9.646663855580764 x^{18} +O(x^{19}).
\end{array}
\end{equation}
Since we are dealing with the  limiting case of perfectly conducting inclusions when the conductivity of inclusions tends to infinity, the effective conductivity is also expected to tend to infinity as a power-law, as the concentration $x$ tends
to the maximal value $x_{c}$ for the hexagonal array,
\begin{equation}
\label{crit}
\sigma(x)\simeq A  (x_c-x)^{-s} .
\end{equation}
The critical superconductivity index (exponent) $s$ believed to be close to $1.3$ \cite{torkbook, perc}. This value is known from numerical simulations, while rigorously it can be anywhere between one and two \cite{gold}. The critical amplitude $A$ is an unknown non-universal parameter.

For regular arrays of cylinders the index is much smaller, $s=\frac 12$ \cite{kell1,McPh} and the critical amplitude is also known with good precision.  Overall effective conductivity of random systems is expected to be higher by order(s) of magnitude as the threshold is approached \cite{LB}.

\section{Critical point}
Probably the simplest way to estimate the position of a critical point, is to apply the diagonal Pade approximants \cite{pade1}, but their direct application leads to poorly convergent, practically random results, with the  best estimate for the threshold  $0.828235$. 
 We attribute the problem  to the trivially ``flat" starting orders in the series (\ref{seriesrand}). In order to compensate for the unchanging values of the coefficients in the starting orders, we consider another sequence of  approximants $F_{n}$ obtained as follows. Let us divide the original series (\ref{seriesrand}) by the function $p(x)=\frac{1-x}{1+x}$ and call the new series $K(x)$. Then
\begin{equation}
F_{n}=p(x) PadeApproximant[K(x),n,n],
\end{equation}
employing again only the diagonal Pade approximants.

There is now a reasonably good sequence of approximations for the critical point, $x_6=0.994313$, $x_{7}=0.978618$, $x_8= 0.822777$, $x_9= 0.882858$.
The percentage error given by the $F_{9}$ equals to  $2.65\%$.

Assuming that $x_c$ is unknown, let us estimate from \eqref{seriesrand} the value of threshold, employing general idea of corrected approximants \cite{corr}. Factor approximations of $\sigma$ can be always represented as a product of two factors: critical part $C(x)=(1-\frac{x}{X_{c}})^{-s}$ and of the rest, i.e. regular part $R(x)$. So one can most generally express the threshold
\begin{equation}
X_{c}=\frac{x  C^{1/s}(x)}{ C^{1/s}(x)-1}.
\end{equation}
The subsequent steps are described below. Suppose we  found explicitly the solution as a factor approximant \cite{fac1,fac2},
\begin{equation} 
\sigma=(2 x+1)^{0.349474}\left(1-\frac{x}{0.93072}\right)^{-1.21092}
\end{equation} 
with approximate threshold value of $x_{0}=0.93072$. Such approximant satisfy the three starting terms from (\ref{seriesrand}) \cite{opt}, and leads to the value of $1.21$ for the index within accepted bounds \cite{gold}. Let us look for another solution in the same form, but with an exact, yet unknown  threshold $X_{c}$,
\begin{equation}
\label{exact}
\sigma'=(2 x+1)^{0.349474}\left(1-\frac{x}{X_c}\right){}^{-1.21092}
\end{equation} 
From here one can express formally, 
\begin{equation}
\label{formal}
X_{c}=\frac{x \left(\frac{\sigma' }{(2x+1)^{0.349474}}\right)^{0.825819}}{\left(\frac{\sigma' }{(2 x+1)^{0.349474}}\right)^{0.825819}-1},
\end{equation} 
since $\sigma'(x)$ is also unknown. All we can do is to use for $\sigma'$ the series (\ref{seriesrand}), so that instead of a true threshold, we have an effective threshold,
\begin{eqnarray}
\label{series1}
X_{c}(x)=&& 0.93072 -4.2551  x^3 +15.125467 x^4-21.963666 x^5...
\nonumber\\
\end{eqnarray}
which should become a true threshold $X_{c}$ as $x\rightarrow X_{c}$! 
Moreover, let us apply re-summation procedure to the expansion 
(\ref{series1})
using again factor approximants $F^{*}(x)$, and define the sought threshold $X_{c}^{*}$ self-consistently, 
\begin{equation}
\label{thr}
X_{c}^{*}=0.93072 -4.2551 x^3 F^{*}(X_{c}^{*}).
\end{equation}
As we approach the threshold, the RHS of (\ref{thr}) should become the threshold. 
 Since factor approximants are defined as $F_{k}^{*}$ for arbitrary number of terms $k$, we will also have a sequence of $X_{c,k}^{*}$. E.g. 
\begin{eqnarray}
\label{thr1}
F_{2}^{*}=(1-0.650467 x)^{5.46479}.
\end{eqnarray}
Expression \eqref{thr1} matches \eqref{thr} up to the 5-order terms included.
 Solving ($\ref{thr}$), we obtain $X_{c,2}^{*}=0.906321$. In the next even order there is no real solution for $X_{c,4}$ and natural stop-sign is generated. 
The percentage error of such estimate is just $0.0638\%$. 

\section{Critical Index $s$}
Conventionally, one would first apply the following transformation,
$
z=\frac{x}{x_c-x}  
$ to the original series, to make calculations with different approximants more convenient.
The most straightforward way to estimate index $s$ is to apply factor approximants \cite{fac1,fac2}  (in terms of the variable $z$), so that possible corrections to the ``mean-field" value unity, appear additively, by definition. Following the standard procedure, the simplest factor approximant is written as follows,
$\sigma_3^{*}=1+b_1 z \left(b_2 z+1\right)^{c_{2}}$,
where $c_2=-0.01357$, $b_1=1.8138$, $b_2=6.8593$, and the critical index $1+c_2= 0.9864$. In the next order the value of critical index improves to $1.0126$. Using even more terms, we obtain

\begin{equation}
\label{f7}
\sigma_7^{*}=1+\frac{b_1 x \left(\frac{b_3 x}{x_c-x}+1\right){}^{c_3} \left(\frac{b_4 x}{x_c-x}+1\right){}^{c_4} \left(\frac{b_2 x}{x_c-x}+1\right){}^{c_2}}{x_c-x},
\end{equation} with $b_1=1.8138$, $b_2=1.0141\, -2.3473 i$, $b_3=1.0141\, +2.3473 i$, $b_4=3.6571$, $c_2=0.0862\, -0.1456 i$, $c_3=0.0862\, +0.1456 i$, $c_4=0.1137$, and the critical index value is good, $s=1+c_2+c_3+c_4=1.28606$. The critical amplitude is equal to $1.55312$.

Let us again transform the original series in terms of $z$, and  to such transformed series $M_{1}(z)$ let us apply the $D-Log$ transformation \cite{pade1,dlog}
and call the transformed series $M(z)$. In terms of $M(z)$ one can readily obtain the sequence of approximations ${s_{n}}$ for the critical index $s$, 
\begin{equation}
s_{n}=\lim_{z\to \infty } (z PadeApproximant[M[z],n,n+1]).\label{seq1}
\end{equation}
Unfortunately, in the case of (\ref{seriesrand}), this method is not accurate. Namely, the best result is $s_{3}= 1.07073$. 
Let us again apply factor approximants, but this time to $M(z)$. The only positive-valued factor approximant appears to be given as follows,
\begin{equation}
\label{betafactor}
\begin{array}{lll}
f_{7}^{*}(z)=A_{0} \left(A_1 z+1\right){}^{c_1} \times
\\
\left(A_2 z+1\right){}^{c_2} \left(A_3 z+1\right){}^{c_3} \left(A_4 z+1\right){}^{-c_1-c_2-c_3-1},
\end{array}
\end{equation}
where $A_{0}=\frac{\pi }{\sqrt{3}}$, $A_1=3.0296\, +1.29635 i$, $A_2=3.0296\, -1.29635 i$, $A_3=0.0614721\, +2.79218 i$, $A_4=0.0614721\, -2.79218 i$, $c_1=-0.707419-0.521627 i$, $c_2=-0.707419+0.521627 i$, $c_3=0.207419\, -0.162683 i$. The critical index is simply,
\begin{equation}
s=A_0 A_1^{c_1} A_2^{c_2} A_3^{c_3} A_4^{-c_1-c_2-c_3-1}=1.29696.
\end{equation} 
Effective conductivity can be reconstructed numerically \cite{puller},
\begin{equation}
\label{numer}
\sigma^{*}(x)=\exp \left(\int_0^{\frac{x}{x_c-x}} f_7^{*}(z)\, dz\right).
\end{equation} Also numerically, the critical amplitude evaluates as $1.37317$. 
Eq. (\ref{numer}) will be compared below with other formula for the effective conductivity valid everywhere.
Let us  look for the solution first in the form of a simple pole,
$f_0(x)=\left(1-\frac{x}{x_{c}}\right)^{-1}$,
so  that our zero approximation $s^{(0)}=1$ for the critical index, is typical for various SCMs.

Let us divide then the original series (\ref{seriesrand}) by $f_{0}$, express the newly found series in term of variable $z$, then apply $D-Log$ transformation and call the transformed series $K(z)$. Finally one can obtain the following sequence of corrected SCM approximations for the critical index,
\begin{equation}
s_{n}=s^{(0)}+\lim_{z\to \infty } (z PadeApproximant[K[z],n,n+1]), \label{cor}
\end{equation}
The ``corrected" sequence of approximate values for the critical index can be calculated readily and we have now three good estimates, $s_0=1.25204$, $s_1=1.24799$ and $s_3=1.38746$. 

Applying different approximants, such as iterated roots \cite{corr}, one can obtain the following sequence of corrected approximations to the critical index,
\begin{equation}
s_{n}=s^{(0)}+\lim_{z\to \infty } (z \;cor_{n}(z)) , \label{corindex}
\end{equation}
where  $cor_{n}(z)$ stands for the iterated root of $n$-th order \cite{corr}, constructed for the series $K(z)$ with such a power at infinity that defines constant correction to $s^{(0)}$.
Calculations with iterated roots are really easy since at each step we need to compute only one new coefficient, while keeping all preceding from previous steps. Namely,
\begin{equation}
\label{seq11}
\begin{array}{lll}
cor_{1}(z)=\frac{0.813799}{3.2288 z+1},
\\
cor_2(z)=\frac{0.813799}{\sqrt{z (12.4821 z+6.45761)+1}},
\end{array}
\end{equation}
and so on iteratively. 
The two starting values 
 for the critical index can be calculated readily, giving $s_1=1.25204$ and $s_2=1.23034$, but in the next orders one obtains complex results.
 In order to continue we define the new series $K_1(z)=\frac{K(z)}{cor_2(z)}$, and apply the technique of iterated approximants to satisfy the new series asymptotically, order-by-order. We can continue the sequence of (\ref{seq1}) (terms $cor_3$ and $cor_4$ are trivial and not shown),
\begin{equation}
\label{seq2}
\begin{array}{lllll}
\frac{cor_5}{cor_2}=1+ \frac{16.7267 z^3}{\left(5.07824 z^2+2.82381 z+1\right)^{3/2}},
\\
\frac{cor_6}{cor_2}= 1+\frac{16.7267 z^3}{6.04206 z^3+\left(5.07824 z^2+2.82381 z+1\right)^{3/2}},
\\
\frac{cor_7}{cor_2}=1+
\\
\frac{16.7267 z^3}{\left(21.1004 z^4+\left(6.04206 z^3+\left(5.07824 z^2+2.82381 z+1\right)^{3/2}\right)^{4/3}\right)^{3/4}},
\end{array}
\end{equation} 
and so on iteratively, so that  $s_5=1.567022$, $s_6=1.450686$, $s_7=1.39582$, $s_8=1.359715$, $s_9=1.347105$, $s_{10}=1.331977$, $s_{11}=1.325325$, $s_{12}=1.31841$, $s_{13}=1.313897$, $s_{14}=1.310147$, $s_{15}=1.306985$, $s_{16}=1.304576$, $s_{17}=1.302397$.
Conversely, the sequence for amplitude is monotonously increasing and ends up in $A_{17}=1.22101$.

Similar techniques were  applied also to the regular case \cite{ourpaper}.
In the random case we proceed by extrapolating from the side of a diluted regime and to the high-concentration regime close to $x_{c}$; while in the regular case we first derived an approximation to the high-concentration regime and then extrapolated to the less concentrated regime. There are indications that physics of a 2D regular and irregular composites is
related to the so-called "necks", certain areas between closely spaced disks \cite{kell1,LB}. 

\section{Final formula  for all concentrations}
We discuss briefly some formulae for the effective conductivity from \cite{method,andr}.
The first formula (Eq.(22), \cite{method}) is nothing else but an improved Pade approximant conditioned by appearance of a simple pole at $x_{c}$.
We also employ Eq.(5) from \cite{andr},  adjusting it with regard to the threshold and critical index values. It exemplifies a crossover from the diluted regime where SCM is valid, to the percolation regime with typical critical behavior.


Closed-form expression for the effective conductivity of the regular hexagonal array of disks is presented in \cite{McPh}.  Since it is defined in the same domain of concentrations as in the random case, a comparison can explicitly quantify the role of a randomness (irregularity) of the composite. But in the most interesting region of large $x$, the relevant formula (Eq 14 from \cite{McPh}) fails. 
 In order to estimate an enhancement factor due to randomness we can still use the numerical results tabulated in \cite{McPh}.  In particular, the enhancement factor at $x=0.9$, is about $15$, compared with (\ref{numer}) and (\ref{3}). The two formulae also happen to be very close to each other everywhere.

Our suggestion for the conductivity valid for all concentrations in the random case  is based on \cite{bender,ourpaper}. Let us apply to $M_1(z)$ another transformation to get 
$T(z)=M_{1}(z)^{-1/s}$, with $s=1.3$, in order to get rid of the power-law behavior at infinity. In terms of $T(z)$ one can readily obtain the sequence of approximations ${A_{n}}$ for the critical amplitude $A$, 
\begin{equation}
A_{n}=x_{c}^s \lim_{z\to \infty } (z PadeApproximant[T[z], n, n + 1])^{-s}.
\end{equation}
There are only few reasonable estimates for the amplitude, $A_0=1.3579$, $A_1=1.05511$, $A_2=1.27363$ and $A_3=1.40546$. 

Following the prescription above, we obtain explicitly,

\begin{equation}
\label{3}
\sigma_{p}^{*}(x)=\frac{0.880698}{\left(\frac{\pi }{\sqrt{12}}-x\right)^{1.3}} 
\left(\frac{6.48521 x^4+9.91426 x^3+5.47416
   x^2+3.32054 x+1}{8.92874 x^3+4.72997
   x^2+2.88473 x+1}\right)^{1.3}.
\end{equation}

\begin{figure}
\begin{center}
\includegraphics[scale=0.6]{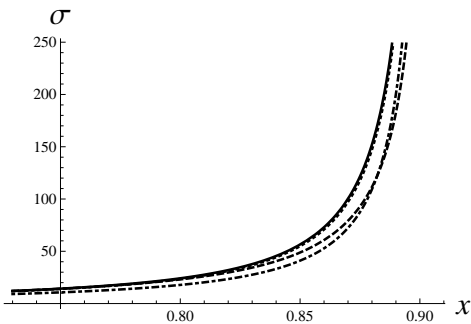}
\end{center}
\caption{
Our formulae (\ref{numer}) (dotted),  (\ref{3})  (solid) are compared with improved Pade approximant from \cite{method} (dashed) and expression from \cite{andr} (dot-dashed).
}
\label{figure}
\end{figure}

Various expressions are shown in Fig.\ref{figure}. Note, that significant deviations of the Pade formula from \cite{method} (with typical value of $s=1$)  compared to our results, start around $x=0.82$. 

\section{Discussion}
In this paper, we developed a direct approach to the effective conductivity of the random 2D arrangements of an ideally conducting cylinders, based on series (\ref{seriesrand}). We confirm the position of a threshold for the effective conductivity, calculate the value of a superconductivity critical index, and obtain a crossover expression \eqref{3}, valid for arbitrary concentrations.  Resummation techniques involved to achieve these goals are original extension of \cite{corr}.  They are in the same mold as the traditional renormalization group \cite{shir,polymer,alt,lett}. 

Our main achievement is a direct (independent on other indices), calculation of the critical index for superconductivity $s=1.3$. Our methods allow thus to correct effectively the value of the critical index given by the large family of  self consistent methods, the most popular among them being ever useful effective medium approximation \cite{chou,Mit2013b}. We cannot yet completely exclude the possibility that $s$ may depend (weekly) on the protocol. Further studies are needed with different protocols.

In a separate paper, we intend to present a generalization of \eqref{seriesrand}, i.e. the  transition formula from the regular hexagonal array to the random array \eqref{seriesrand}. 
We expect to obtain a dependence of the critical index on the degree of randomness. 

\section*{Acknowledgement} We are grateful to Wojciech Nawalaniec for computer derivation of the formula \eqref{seriesrand}.


\begin{thebibliography}{99}

\bibitem{torkbook}
S.Torquato, Random Heterogeneous Materials: Microstructure and Macroscopic Properties, Springer-Verlag. New York, 2002

\bibitem{Mit2013b}
V. Mityushev, N.Rylko,
Maxwell's approach to effective conductivity and its limitations, 
The Quarterly Journal of Mechanics and Applied Mathematics (2013); doi: 10.1093/qjmam/hbt003

\bibitem{torkreview} S.Torquato, F. H. Stillinger, Jammed hard-particle packings: From Kepler to Bernal and beyond, Reviews of Modern Physica, {\bf 82}, 2634 (2010), 

\bibitem{method}
R.Czapla, W.Nawalaniec and V.Mityushev,
Effective conductivity of random two-dimensional composites with circular non-overlapping inclusions, 
Comput. Mat. Sci.{\bf 63}, 118 (2012)

\bibitem{ourpaper} 
S. Gluzman, V. Mityushev, W. Nawalaniec,
Crossover in the effective conductivity of the regular array of ideal conductors 2014, in press

\bibitem{protoc1}
T. Quickenden, G.K. Tan, 
Random packing in two dimensions and the structure of monolayers
Journal of Colloid and Interface Science, {\bf48}, 382-(1974)

\bibitem{protoc2}
B. Lubachevsky, F.H. Stillinger, 
Geometric properties of random disk packings,
J. Stat. Phys.{\bf 60},  561 (1990)


\bibitem{protoc3}
R. Jullien,  J. F. Sadoc  and  R. Mosseri, 
Packing at Random in Curved Space and Frustration: a Numerical Study, 
J. Phys. France {\bf 7} ,1677, (1997) DOI: 10.1051/jp1:1997162 


\bibitem{CzNawMit2}
R.Czapla, W. Nawalaniec and V. Mityushev, 
Simulation of representative volume elements for random 2D composites with circular non-overlapping inclusions, 
Theoretical and Applied Informatics, {\bf 24}, 227 (2012)

\bibitem{berr} J.G.Berryman, 
Random close packing of hard spheres and disks. 
Phys.Rev.A, {\bf 27}, 1053(1983)

\bibitem{chou} Tuck C.Choy, 
Effective Medium Theory. Principles and Applications. Clarendon Press. Oxford. 1999

\bibitem{kell0} J.B. Keller, 
A Theorem on the Conductivity of a Composite Medium 
 J. Math. Phys. {\bf 5}, 548 (1964)

\bibitem{ryl}  N. Rylko, 
Transport properties of the regular array of highly
conducting cylinders. 
{\it J, Engrg. Math} {\bf 38}, 1  (2000).

\bibitem{ryl2008}  N. Rylko, 
Structure of the scalar field around
unidirectional circular cylinders. 
Proc. R. Soc. A{\bf 464}, 391 (2008)


\bibitem{fix}  V.I. Yukalov and S. Gluzman, 
Critical Indices as Limits of Control Functions 
Phys. Rev. Lett. {\bf 79}, 333 (1997)



\bibitem{low}
 J. Adler, Y. Meir, A. Aharony,
A.B. Harris and L. Klein, 
Low-Concentration Series in General Dimension,
 J. Stat. Phys. {\bf 58}, 511 (1990)



\bibitem{perc}  J.P. Clerc, G. Giraud, J.M. Laugier and J.M. Luck, 
The electrical conductivity of binary
disordered systems, percolation clusters, fractals and related models. 
Adv.Phys. {\bf 39}, 191 (1990)

\bibitem{gold} K.Golden, Phys.Rev.Lett. {\bf 65}, 2923
Convexity and Exponent Inequalities for Conduction near Percolation 
 (1990)

\bibitem{kell1} J.B. Keller, 
Conductivity of a Medium Containing a Dense Array of Perfectly Conducting Spheres or Cylinders or Nonconducting Cylinders, 
J. Appl. Phys. {\bf 34},  991  (1963)

\bibitem{McPh} 
W.T. Perrins, D.R. McKenzie and R.C. McPhedran,
Transport properties of regular array of cylinders. 
Proc.R. Soc.A {\bf 369}, 207 (1979)

\bibitem{LB} 
L. Berlyand, A. Novikov
Error of the network approximation for densely packed composites with irregular geometry, 
SIAM J. Math. Analysis, {\bf 34(2)}, 385 (2002)



\bibitem{pade1}  G.A. Baker and P. Graves-Moris, 
Pad\'e Approximants (Cambridge
University, Cambridge, 1996)

\bibitem{corr} 
S.Gluzman, V.I.Yukalov, 
Self-similar extrapolation from weak to strong coupling,
 J.Math.Chem.{\bf  48}, 883 (2010)

\bibitem{fac1}  S. Gluzman, V.  I. Yukalov, D. Sornette, 
Self-Similar Factor Approximants, 
Phys. Rev. E {\bf 67 (2)}, art. 026109 (2003)

\bibitem{fac2} V. I. Yukalov, S. Gluzman, D. Sornette, 
Summation of Power Series by Self-Similar Factor Approximants, 
Physica A, {\bf 328}, 409 (2003) 

\bibitem{opt} V. I. Yukalov and S. Gluzman, 
Optimization of Self-Similar Factor Approximants,
 Molecular Physics, 
 {\bf 107}, 2237 (2009)

\bibitem{dlog}
S. Gluzman and V.I. Yukalov, European Journal of Applied Mathematics (2014), to appear

\bibitem{puller} S. Gluzman, D.A. Karpeev, L.V. Berlyand,
Effective viscosity of
puller-like microswimmers: a renormalization approach.
 J. R. Soc. Interface {\bf 10}: 20130720 (2013)

\bibitem{andr}
I.V. Andrianov, V.V. Danishevskyy, A. L. Kalamkarov, 
Analysis of the effective conductivity of composite materials in the entire range of volume fractions of inclusions up to the percolation threshold, 
Composites: Part B {\bf 41}, 503 (2010) 

\bibitem{bender} C.M. Bender and S.Boettcher,
Determination of $f(\infty)$ from the asymptotic series for $f(x)$ about $x=0$ 
 J. Math.Phys. {\bf 35}, 1914 (1994)


\bibitem{shir} D.V. Shirkov,
The renormalization group, the invariance principle, and functional self-similarity.
Sov. Phys. Dokl. {\bf 27}, 197 (1982)

\bibitem{polymer}
J. Des Cloizeaux, R. Conte, R. and G. Jannink, 
Swelling of an isolated polymer chain in a solvent, 
J. Physique Lett. {\bf 46}, L595 (1985)

\bibitem{alt}
A.R. Altenberger, J.S. Dahler.
A renormalization
group calculation of the viscosity of a hard-sphere
suspension.
 J. Colloid Interface Sci. {\bf 189}, 379 (1997)
(doi:10.1006/jcis.1997.4849)

\bibitem{lett}
S. Gluzman, V.I. Yukalov, 
Self-similar continued root approximants,
 Phys. Lett. A {\bf 377}, 124 (2012) 
 
%
%


\end{thebibliography}
\end{document}